\documentclass[sigconf, 10pt, nonacm]{acmart}
\title{Can You Check That? The Checkability Boundary for Local LLM Network Automation\vspace{0.2in}}

\author{Maleeha Masood}
\affiliation{
  \institution{University of Illinois Urbana-Champaign}
  \country{Urbana, IL, USA}
}
\author{Momina Nofal}
\affiliation{
\institution{Independent Researcher}
\country{USA}
}

\newcommand{\name}{Touchstone}

\usepackage{listings}
\usepackage{xcolor}
\usepackage{tabularx}

\usepackage{tikz}
\usetikzlibrary{arrows.meta,positioning,calc,fit,backgrounds}
\usepackage[normalem]{ulem}

\usepackage[most]{tcolorbox}

\newtcolorbox{definitionbox}{
  colback=magenta!8,
  colframe=black,
  title=Definition: Intrinsic Check,
  fonttitle=\bfseries,
  boxrule=0.6pt,
  arc=2mm
}

\lstdefinestyle{reqstyle}{
  basicstyle=\ttfamily\scriptsize,
  breaklines=true,
  columns=fullflexible,
  frame=none,
  aboveskip=2pt, belowskip=2pt
}

\definecolor{jsonkey}{RGB}{20,80,160}
\definecolor{jsonstr}{RGB}{160,60,20}
\lstdefinestyle{jsonstyle}{
  basicstyle=\ttfamily\scriptsize,
  breaklines=true,
  columns=fullflexible,
  frame=none,
  showstringspaces=false,
  stringstyle=\color{jsonstr},
  aboveskip=2pt, belowskip=2pt
}

\newcommand{\squishlist}
{
    \begin{list}{$\bullet$}
    {
        \setlength{\itemsep}{0pt}      \setlength{\parsep}{2pt}
        \setlength{\topsep}{2pt}       \setlength{\partopsep}{0pt}
        \setlength{\leftmargin}{1em} \setlength{\labelwidth}{0.5em}
        \setlength{\labelsep}{0.5em}
    }
}
\newcommand{\squishend}
{
    \end{list}
}

\begin{abstract}
Sending every network-automation input to a third-party frontier LLM exports sensitive artifacts such as production configurations, topologies, and logs, and incurs substantive costs. Querying small language models (SLMs) locally avoids these concerns, but SLM outputs can be too error-prone for direct use. This work introduces \emph{checkability} as a criterion for determining which tasks are suitable for local inference. A task is checkable when it exposes a cheap, deterministic test - an \emph{intrinsic check} - that rejects outputs violating a necessary correctness condition. 
We instantiate this idea in \emph{\name{}}, a local-first pipeline that uses seven off-the-shelf SLMs (1--8B parameters) to generate candidates, uses task-specific intrinsic checks to reject responses, and escalates unresolved inputs to a frontier LLM. On conflict detection and intent translation tasks, \name{} reaches 98.6\% and 93.8\% end-to-end accuracy while escalating only 16\% and 17\% of inputs, respectively. On TeleQnA, a knowledge-only control that has no task-specific intrinsic checks, \name{} is unable to match the accuracy of the frontier baseline. Our results support a simple deployment rule: keep inference local when task semantics support precise, low-cost checks; escalate the rest.
\end{abstract}

\begin{document}

\maketitle

\section{Introduction}

Network operators are unlikely to rely entirely on a model they cannot expose their data to, cannot inspect, or simply, cannot afford to run. This is the uncomfortable gap in today’s LLM-driven network automation. Frontier-scale models have made ambitious tasks like translating operator intent, triaging incidents, detecting anomalies, generating configurations, and synthesizing control code more feasible \cite{vpp,netconfeval,llm-network-code,netllm,netintent,agenticimr,polysmith,nada}. Yet, LLM inference often requires exporting sensitive operational artifacts to third-parties, while incurring substantial inference costs. On-premise inference can avoid this sharing of data, but running an LLM requires significant GPU infrastructure and operational expertise. This leaves network automation in a paradoxical state where the strongest models are often least usable where their help would matter most.

Small language models (SLMs) i.e. language models with $\leq$8B parameters, appear to close this gap. An SLM can run inside an operator's environment, keep sensitive data within the network boundary, and avoid the cost and governance burden of remote frontier inference \cite{ibm_slm}. However, SLMs are known to be have weaker reasoning, are more prone to hallucinations and have smaller context windows than LLMs \cite{netconfeval, vpp, onioneval}. Thus, treating an SLM as merely a smaller frontier LLM is the wrong abstraction. SLMs can only be safely deployed for network automation when there is a principled rule for admitting local outputs and escalating the remainder to a frontier LLM.

We introduce \textit{intrinsic checks} to decide if a response from an SLM is safe to use. An intrinsic check is a task-specific, deterministic predicate over the task input, a candidate output, and optional network state that tests a necessary condition for correctness. 

Many network automation tasks expose such intrinsic checks. A reachability specification should not both permit and deny the same location--prefix pair. An intent translation should not introduce locations or prefixes absent from the request. A log template should regenerate the concrete log line it summarizes. Routing code can be executed against tests on concrete topologies. Intrinsic checks are simple, inexpensive checks and do not compare against a reference answer, or invoke a verifier or LLM.

We instantiate \textit{intrinsic checks} in \textit{\name{}}\footnote{A touchstone is a test or criterion for determining the quality or genuineness of a thing. Historically, it is a stone used to test the purity of precious metals.}, a local-first pipeline that uses seven heterogeneous, off-the-shelf SLMs with 1--8B parameters. For each input, the SLMs generate candidates, and task-specific intrinsic checks reject candidates that violate necessary conditions. \name{} then selects from remaining candidates or escalates input when no candidate passes the intrinsic checks. In \name{}, the SLMs provide diversity, intrinsic checks determine eligibility for accepting an SLM's response, and frontier escalation handles the unresolved residue. Instead of completely eliminating third-party inference, \name{} just makes it selective.

We evaluate \name{} on 4 different structured networking tasks, namely conflict detection, intent translation, log parsing, and generating routing code, and a knowledge based dataset on telecommunications called TeleQnA. We report end-to-end accuracy and frontier LLM escalation rate, and compare against a baseline that sends every input to the frontier LLM. \name{} reaches 98.6\% end-to-end accuracy on conflict detection with 16\% escalation and zero false accepts, and 93.8\% accuracy on intent translation with 17\% escalation. On TeleQnA, a knowledge-only control for which there are no task-specific intrinsic checks, \name{} remains below the frontier baseline. This negative result is instructive: model diversity and selective escalation alone cannot match all-frontier performance unless the task exposes intrinsic checks. Together, these results support using intrinsic checks as a practical deployment boundary for local-first LLM network automation.

This paper makes three contributions:

\squishlist

    \item We identify \textit{checkability} as the boundary for local deployment: tasks whose outputs can be cheaply and deterministically checked should remain local, while tasks without such checks should be escalated to a frontier LLM.
    
    \item We define \textit{intrinsic checks} as deterministic tests that reject a candidate when it makes a claim the task input or network state refutes.

    \item We evaluate \name{} on four structured networking tasks and a knowledge-only control, quantifying the end-to-end accuracy--escalation tradeoff.

\squishend
\section{Intrinsic Checks}
\label{sec:checks}

\vspace{0.05in}
\begin{definitionbox}
Let (P(x,y)) mean “$y$ is correct for input $x$.” An \emph{intrinsic check} is a deterministic predicate $Q(x, y, s)$, with optional network state $s$, such that correctness implies the check:
\[
P(x, y) \implies Q(x, y, s).
\]
That is, $Q$ is a \emph{necessary} condition for correctness, but not a sufficient one.

An intrinsic check uses no ground truth, LLM, learned verifier, or remote API. Since \(Q\) is only necessary, it is one-sided:

\squishlist
    \item A \textbf{failed} check ($\neg Q$) entails $\neg P$: the candidate is refuted, because it violates a condition every correct answer must satisfy.
    \item A \textbf{passed} check ($Q$) only means the candidate survived refutation. Its reliability depends on the \emph{false-accept rate} (FAR)—the fraction of accepted candidates that are wrong—which must be measured.
\squishend
\end{definitionbox}

Many network automation tasks produce artifacts with constraints that any correct answer must satisfy. These constraints for correctness can be tested for with cheap deterministic tests that we call \emph{intrinsic checks}.

We use existing networking datasets to generate task-specific intrinsic checks.

\textbf{Intent translation.}
In NetConfEval formal-specification translation~\cite{netconfeval}, the task is to compile natural-language requirements into a JSON specification containing reachability, waypoint, and load-balancing rules. The simplest syntactic check is that the output must be a dictionary with the three expected keys. The intrinsic checks however go beyond the shape of the object and ask whether its contents are supported by the request. For example, the input

\begin{center}
\begin{minipage}{0.9\linewidth}
\begin{lstlisting}[style=reqstyle, frame=single, framerule=0.4pt, framesep=4pt]
Traffic originating from barcelona can reach the subnet 100.0.9.0/24.
\end{lstlisting}
\end{minipage}
\end{center}

should produce a specification such as:

\begin{center}
\begin{minipage}{0.9\linewidth}
\begin{lstlisting}[style=jsonstyle, frame=single, framerule=0.4pt, framesep=4pt]
    {"loadbalancing": {}, 
    "reachability": {"barcelona": ["100.0.9.0/24"]}, 
    "waypoint": {}}
\end{lstlisting}
\end{minipage}
\end{center}

We develop two intrinsic checks for this task. First, the output should be \emph{grounded}: every prefix and location in the generated specification should appear in the input request. Second, the output should provide \emph{coverage}: the number of generated rule atoms should roughly match the number of requirement clauses in the request. 
These checks reject common SLM failures such as hallucinating unseen subnets, or silently dropping large portions of the request. However, they do not prove that a grounded and complete-looking specification correctly captures the operator’s intent.

\textbf{Conflict detection.}
In NetConfEval translation conflict detection~\cite{netconfeval}, the task is to decide whether a set of natural-language requirements contains a contradiction. For e.g.:

\begin{center}
\begin{minipage}{0.9\linewidth}
\begin{lstlisting}[style=reqstyle, frame=single, framerule=0.4pt, framesep=4pt]
The subnet 100.0.22.0/24 is reachable from amsterdam.
geneva cannot reach 100.0.4.0/24.
100.0.4.0/24 is accessible from geneva.
\end{lstlisting}
\end{minipage}
\end{center}

The correct output is:
CONFLICT

The request is conflicting because the same $\langle$location, prefix$\rangle$ pair is asserted both reachable and unreachable. The intrinsic check for this task extracts reachability predicates and tests \emph{consistency}. Specifically, it parses each requirement sentence into $\langle$location, prefix$\rangle$ pairs, classifies each as reachable or unreachable using the text in the input, and flags a conflict whenever the same pair is asserted both ways, deterministically catching contradictions. If the input contains a conflict, any candidate that says “no conflict” is rejected. If no contradiction is extracted, the check does not prove the request is conflict-free.

\textbf{Log parsing.}
In Loghub's OpenSSH, HDFS and Proxifier datasets \cite{loghub}, the task is to convert a raw log line into its event template, replacing every variable value (IPs, IDs, hostnames, ports) with a wildcard \texttt{<*>}. For example:

\begin{center}
\begin{minipage}{0.9\linewidth}
\begin{lstlisting}[style=reqstyle, frame=single, framerule=0.4pt, framesep=4pt]
Failed password for invalid user admin from 5.188.10.180 port 60682 ssh2
\end{lstlisting}
\end{minipage}
\end{center}

should produce a template such as:

\begin{center}
\begin{minipage}{0.9\linewidth}
\begin{lstlisting}[style=jsonstyle, frame=single, framerule=0.4pt, framesep=4pt]
Failed password for invalid user <*> from <*> port <*> ssh2
\end{lstlisting}
\end{minipage}
\end{center}

The intrinsic check we develop is invertibility: the generated template must be able to regenerate the original log line by replacing wildcards with concrete substrings. 

This check is a valid Q, but with a larger false-accept rate, since many over-general templates can still regenerate the line. 

\textbf{Routing code generation.}
In NetConfEval's Routing Code Generation task~\cite{netconfeval}, the model must write a Python function that returns the routing paths between hosts. For example, for the shortest-path policy and a topology in which hosts \texttt{h1} and \texttt{h2} share a LAN \texttt{A}, the function must satisfy:

\begin{center}
\begin{minipage}{0.9\linewidth}
\begin{lstlisting}[style=reqstyle, frame=single, framerule=0.4pt, framesep=4pt]
route({'h1': {0: 'A'}, 'h2': {0: 'A'}})
== {'h1': {'h2': ['h1', 'h2']},
'h2': {'h1': ['h2', 'h1']}}
\end{lstlisting}
\end{minipage}
\end{center}

    and the model should produce a function such as: 

\begin{center}
\begin{minipage}{0.9\linewidth}
\begin{lstlisting}[style=jsonstyle, frame=single, framerule=0.4pt, framesep=4pt]
def route(topology, requirements=None):
lan = {}
for d, ports in topology.items():
for l in ports.values():
lan.setdefault(l, []).append(d)
adj = {d: {n for l in p.values()
for n in lan[l] if n != d}
for d, p in topology.items()}
# BFS shortest paths between host pairs
...
return paths
\end{lstlisting}
\end{minipage}
\end{center}

The intrinsic check here is \emph{execution}: the generated function is run against the task's test cases. Unlike the checks above, execution-based checking is already standard practice for code. We include it as the strongest example of intrinsic checking

Intrinsic checks are cheap to execute because they compare a candidate’s claims against facts or constraints already available in the input or network state using ordinary deterministic operations. Each check defines an admission boundary: candidates for which no error witness is found may be accepted locally, while the remaining candidates are escalated. However, passing a check does not prove that a candidate is correct. An incorrect candidate may pass when its error leaves no detectable witness. We therefore characterize each check by its false-accept rate, measured separately for each task in \S\ref{sec:eval-accuracy}.

\paragraph{Uncheckable Tasks} A useful intrinsic check works because passing it rules out a class of wrong answers. A passing candidate has survived a necessary condition, so it is less likely to be wrong. This depends on the task exposing a property that correct and incorrect answers do not share. We find that knowledge-based tasks such as TeleQnA \cite{telqna}, which tests telecommunications knowledge, expose no such property. Any answer choice is a well-formed, valid option, but it can still be wrong, because its correctness depends entirely on external knowledge rather than task-intrinsic structure.

\paragraph{Who Writes the Checks?}
Intrinsic checks are written once per task by the system designer, rather than per query by the operator. In our prototypes, these checks are small: the grounding-and-coverage check for intent translation requires 49 lines of code, the check for conflict detection requires 43, and the invertibility check for log parsing requires 28. The designer does not need to solve the task or specify every valid output. Instead, the designer identifies necessary conditions that any valid output must satisfy, such as using only prefixes present in the request or preserving required waypoints. Each condition must be a sound refuter: violating it must provide valid grounds for rejecting the candidate. The resulting check is then evaluated on a development set using two quantities: its false-accept rate, which captures incorrect candidates that pass, and its unnecessary-escalation rate, which captures correct candidates that are rejected. These measurements make the check’s residual risk and loss of local coverage explicit before deployment.

\section{The \name\ Workflow}
\label{sec:workflow}

\begin{figure}[t]
\centering
\resizebox{\columnwidth}{!}{%
\begin{tikzpicture}[
    box/.style={
        rectangle,
        rounded corners=5pt,
        draw=black!70,
        align=center,
        minimum height=8mm,
        text width=2.45cm,
        font=\small,
        inner sep=4pt
    },
    widebox/.style={
        rectangle,
        rounded corners=5pt,
        draw=black!70,
        align=center,
        minimum height=8mm,
        text width=2.85cm,
        font=\small,
        inner sep=4pt
    },
    gate/.style={
        rectangle,
        draw=black!70,
        align=center,
        minimum width=2.05cm,
        minimum height=8mm,
        inner sep=2pt,
        font=\small
    },
    phaselabel/.style={
        font=\bfseries\small,
        anchor=west,
        align=left
    },
    arrow/.style={
        -{Stealth[length=2mm]},
        thick,
        draw=black!80
    },
    dottedarrow/.style={
        -{Stealth[length=2mm]},
        thick,
        dotted,
        draw=black!75
    }
]


\draw[
    fill=blue!14,
    draw=black!65,
    rounded corners=6pt
] (-3.20,1.90) rectangle (1.30,-0.05);

\draw[
    fill=orange!17,
    draw=black!65,
    rounded corners=6pt
] (-3.20,-0.22) rectangle (1.30,-7.00);

\draw[
    fill=teal!18,
    draw=black!65,
    rounded corners=6pt
] (1.43,0.95) rectangle (6.75,-7.00);


\node[phaselabel] at (-2.32,1.48)
{(1) Offline Profiling};

\node[phaselabel] at (-2.28,-6.38)
{(2) Local Generation\\\& Aggregation};

\node[phaselabel] at (2.68,-6.45)
{(3) Confidence-\\Gated Escalation};


\node[
    box,
    fill=gray!10,
    text width=2.15cm
] (weights) at (-0.85,0.68)
{\textbf{Task weights}};


\node[
    box,
    fill=blue!10
] (slms) at (-0.85,-1.13)
{\textbf{7 local SLMs}\\candidate artifacts};

\node[
    widebox,
    fill=green!12
] (agg) at (-0.85,-3.17)
{\textbf{Aggregation}\\
 \textbf{+ Intrinsic Checks}\\
 task-specific};

\node[
    box,
    fill=blue!10
] (candidate) at (-0.85,-5.25)
{\textbf{Aggregated}\\
 \textbf{local candidate}\\
 $\hat{a}$};


\node[
    box,
    fill=blue!10,
    text width=1.95cm
] (final) at (4.10,0.08)
{\textbf{Final}\\answer};

\node[
    box,
    fill=green!15,
    text width=1.90cm
] (local) at (2.65,-1.28)
{\textbf{Return}\\
 \textbf{local} $\hat{a}$};

\node[
    box,
    fill=red!15,
    text width=1.90cm
] (frontier) at (5.55,-1.28)
{\textbf{Frontier}\\
 \textbf{Escalation}};

\node[
    gate,
    fill=orange!15
] (gate) at (4.10,-3.15)
{$\mathrm{conf}(x)<\tau$?\\[-0.5mm]
 $\tau=0.6$};

\node[
    widebox,
    fill=green!10
] (conf) at (4.10,-5.22)
{\textbf{Confidence}\\
 $\mathrm{conf}(x)$ from\\
 $\mathrm{agree}(\hat{a})$, adjusted\\
 by intrinsic check};


\draw[arrow] (slms.south) -- (agg.north);

\draw[arrow] (agg.south) -- (candidate.north);

\draw[dottedarrow]
    (weights.west)
    to[out=205,in=155]
    node[
        font=\footnotesize,
        fill=orange!17,
        inner sep=1pt,
        pos=0.56
    ] {$w_m$}
    (agg.west);


\draw[arrow]
    (candidate.east) --
    (conf.west);

\draw[arrow] (conf.north) -- (gate.south);



\coordinate (finalleft) at
    ($(final.south west)!0.28!(final.south east)$);

\coordinate (finalright) at
    ($(final.south west)!0.72!(final.south east)$);

\draw[arrow]
    (local.north)
    -- ++(0,0.25)
    -| (finalleft);

\draw[arrow]
    (frontier.north)
    -- ++(0,0.25)
    -| (finalright);

\draw[arrow] (gate.north west) -- (local.south);
\draw[arrow] (gate.north east) -- (frontier.south);

\node[font=\footnotesize] at (3.28,-2.35) {no};
\node[font=\footnotesize] at (4.92,-2.35) {yes};

\end{tikzpicture}%
}
\caption{\name{}'s local-first workflow}
\label{fig:overview}
\end{figure}
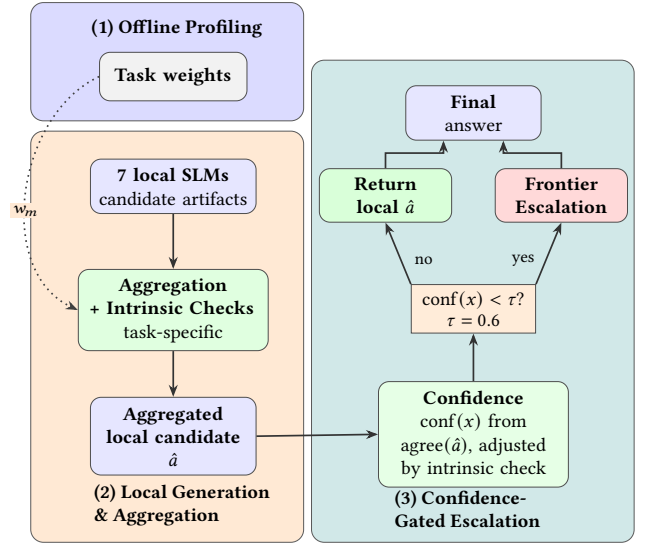

\name{} is an admission-control architecture for local SLM inference with intrinsic checks. We say that a candidate is \textit{admitted} when the system accepts it without escalating the input to a frontier LLM. \name{} works by separating generation from admission: a family of SLMs proposes candidate answers, and intrinsic checks reject candidates that violate necessary conditions. A confidence score then decides if the local answer is admissible or if the query needs to be escalated to a frontier LLM to resolve.

\name{} runs seven off-the-shelf SLMs---Llama-1B, Qwen-1.5B, SmolLM-1.7B, Qwen-3B, Llama-3B, Qwen-7B, and Llama-8B---in three phases: (1) offline profiling, (2) local generation and aggregation, and (3) confidence-gated escalation.

\textbf{Offline profiling.}
\name\ profiles each SLM once on a small held-out development split. This task-specific accuracy is then used to determine the model's competence on the task and consequently, its voting weight $w_m$. Our development sets contain 23.1--25.0\% of the available data i.e. 100-150 examples for each task.

For each model $m$, \name\ computes
\[
  w_m = \max\!\bigl(0.02,\; \mathrm{acc}_m - \mathrm{chance}\bigr)^{1.5},
\]
where $\mathrm{acc}_m$ is the model's accuracy on the corresponding development split and $\mathrm{chance}$ is the random baseline ($0.5$ for conflict detection, $0$ for the open-ended generation tasks). 
Subtracting chance prevents near-random models from dominating the aggregate, and the exponent emphasizes models that are reliably above chance.

\textbf{Local generation and aggregation.}
For each input, \name\ queries all seven SLMs and aggregates their outputs using a task-specific procedure. The aggregation method depends on the structure of the output: conflict detection and log parsing produce a single answer, intent translation produces a set of rules, and routing-code generation may admit multiple correct programs.

For conflict detection and log parsing, \name\ uses a \emph{skill-weighted vote}. Each model \(m\) produces a prediction with weight \(w_m\). For every distinct prediction, \name\ sums the weights of the models that produced it and selects the prediction with the largest total weight. The intrinsic check constrains this selection. For conflict detection, a consistency check overrides the vote when it establishes a contradiction. For log parsing, an invertibility check excludes templates that cannot regenerate the original log line. 

Routing-code generation requires no voting because several implementations may be correct. Instead, \name\ executes every generated function against the test suite and retains each candidate that passes the execution check.

Intent translation requires finer-grained aggregation because each output is a set of rules, and a model may generate some rules correctly while missing or misgenerating others. \name\ therefore decomposes every generated specification into \emph{atoms}, with one atom per rule. For example, the rule that Paris must reach prefix \texttt{100.0.1.0/24} becomes
\texttt{("reach", "paris", "100.0.1.0/24")}.
\name\ pools atoms across all seven models, merges identical atoms, and retains an atom when the models that generated it account for at least half of the ensemble's total weight. Thus, an atom may be retained through agreement among several models or through the support of one highly weighted model. This rule-level aggregation can construct a correct specification from complementary outputs even when no individual model produces the entire specification correctly. However, atom-wise recombination can also introduce an incoherent combination of individually plausible rules. To guard against this failure, \name\ compares the assembled specification with the individual specification that scored highest on the intrinsic checks and retains whichever of the two that receives the higher grounding-and-coverage score.

At the end of this phase, \name\ produces one aggregated candidate for conflict detection, log parsing, and intent translation. For routing-code generation, it retains all candidates that pass the execution check.

\textbf{Confidence-gated escalation.}
After aggregation, \name\ assigns the selected local candidate a confidence score
\(\mathrm{conf}(x)\in[0,1]\). This score combines two signals: the weighted support of the SLM family and the evidence supplied by the task's intrinsic check. \name\ returns the local candidate when its confidence meets a threshold \(\tau\) and otherwise escalates to a frontier LLM:
\[
    \mathrm{escalate}(x)
    \quad\Longleftrightarrow\quad
    \mathrm{conf}(x) < \tau,
    \qquad \tau = 0.6.
\]
Thus, escalation is a single per-input decision made only after aggregation.

The base confidence is the \emph{family agreement}: the fraction of the ensemble's total skill weight supporting the selected candidate. For tasks with a single-valued output,
\[
    \mathrm{agree}(\hat{a})
    =
    \frac{\sum_{m:\,\mathrm{pred}_m=\hat{a}} w_m}
         {\sum_m w_m},
\]
where \(\hat{a}\) is the aggregated prediction. For intent translation, whose output is assembled from multiple rule atoms, \name\ instead uses the mean weighted support of the atoms in the assembled specification.

The intrinsic check then modifies this base confidence. The exact rule is task-specific because the checks provide different strengths of evidence.

For log parsing, \name\ first excludes templates that fail the invertibility check and uses the family agreement of the selected passing template as its confidence. For intent translation, the grounding-and-coverage check identifies concrete defects such as entities not grounded in the request or missing required rules. Such defects discount the agreement score and can push an otherwise well-supported specification below the escalation threshold.

For conflict detection, the check is one-sided: it can prove that two policies conflict, but it cannot prove that they are consistent. When the check establishes a contradiction, \name\ selects \texttt{CONFLICT} and sets
\(\mathrm{conf}(x)=1\). Otherwise, confidence remains the weighted agreement behind the selected label. For routing code, execution provides a stronger acceptance signal than model agreement. \name\ therefore sets confidence to \(1\) when at least one generated function passes the test suite and to \(0\) when none does.

When \name\ escalates an input, it includes the check's diagnostic evidence---for example, an unreproduced token, an ungrounded entity, or a failed execution test---in the frontier-model prompt as a repair hint. \name\ returns the frontier model's reponse as the final output.

\section{Evaluation}
\label{eval}
\begin{table}[t]
\centering
\small
\setlength{\tabcolsep}{2.5pt}
\begin{tabular}{lrrrrr}
\toprule
Dataset & $n$ & Best SLM & GPT-5.5 & \name{} & Egress \\
\midrule
\multicolumn{6}{l}{\emph{Structured Tasks}}\\
Conflict det.     & 500 & 0.840 & 1.000 & 0.986 & 16\% \\
Intent transl.    & 500 & 0.740 & 0.992 & 0.938 & 17\% \\
Log parse (SSH)   & 300 & 0.743 & 0.997 & 0.997 & 37\% \\
Log parse (HDFS)  & 300 & 0.863 & 0.987 & 0.917 &  7\% \\
Log parse (Prox.) & 300 & 0.650 & 0.917 & 0.877 & 42\% \\
Routing code      &   8 & 0.125 & 0.500 & 0.500 & 75\% \\
\addlinespace
\multicolumn{6}{l}{\emph{Knowledge Tasks}}\\
TeleQnA           & 500 & 0.780 & 0.842 & 0.782 & 16\% \\
\bottomrule
\end{tabular}
\caption{\name{} accuracy and frontier egress across all tasks, compared with GPT-5.5 and Best Single SLM accuracy.}
\label{tab:results}
\end{table}

{\setlength{\textfloatsep}{6pt plus 2pt minus 2pt}
\begin{table}[t]
\centering
\small
\begin{tabular}{llr}
\toprule
Check & Task & FAR \\
\midrule
Consistency & Conflict detection & 0.0\% \\
Grounding + coverage & Intent translation & 5.6\% \\
Invertibility        & Log (OpenSSH)       & 27.7\% \\
Invertibility        & Log (HDFS)          & 14.4\% \\
Invertibility        & Log (Proxifier)     & 30.0\% \\
\bottomrule
\end{tabular}
\caption{False-accept rate (FAR) of each intrinsic check on the evaluation split.}
\label{tab:far}
\end{table}}

We evaluate \name{} based on its accuracy, egress (the percentage of input getting escalated to the frontier LLM, GPT-5.5), and the false-accept rates of the intrinsic checks of each task.

We use four existing structured networking task datasets: conflict detection, intent
translation, log parsing, and routing-code generation. These cover logical,
structural, invertibility-based, and execution-based intrinsic checks (\S\ref{sec:checks}). We also
include TeleQnA~\cite{telqna} as a knowledge-only control which tests the boundary case where \name{} has no intrinsic
check to apply.

Table~\ref{tab:results} reports accuracy and egress against an all-frontier baseline. Table~\ref{tab:far} reports false-accept rate (FAR) of each task's checks: among items the check accepted locally, what fraction were actually wrong. This isolates check reliability.

\subsection{Accuracy, Egress and False-accept Rates}
\label{sec:eval-accuracy}

\name{} performs best on the two tasks with the lowest false-accept rates. On conflict detection, it reaches 98.6\% accuracy, compared with 100\% for GPT-5.5, while escalating only 16\% of inputs. We observe no false accepts: when the input contains an explicit contradiction, the check rejects any candidate that incorrectly reports no conflict. On intent translation, \name{} reaches 93.8\% accuracy while escalating 17\% of requests, compared with 99.2\% accuracy when every request is sent to GPT-5.5. Its 5.6\% false-accept rate reflects the limits of the grounding and coverage checks: they catch hallucinated entities and omitted clauses, but do not prove that the generated policy is semantically correct.

Log parsing shows the limits of a weak check. Invertibility is necessary, but not very selective: an over-general template can still regenerate the original log line and therefore pass. As a result, the false-accept rate is high—27.7\% on OpenSSH and 30.0\% on Proxifier i.e. nearly one-third of templates that pass the check are still incorrect. \name{} compensates by escalating 37\% and 42\% of inputs, respectively. Its accuracy therefore comes largely from frontier fallback rather than from the check itself, which is too permissive to serve as a reliable admission signal.

HDFS illustrates a different failure mode. It escalates only 7\% of inputs—the lowest rate among the structured tasks—but reaches 91.7\% accuracy, compared with 98.7\% for the frontier model. Its 14.4\% false-accept rate explains the gap: low egress is useful only when locally admitted outputs are reliable. Keeping incorrect answers local is not a success, even if it reduces frontier use.

Routing code and TeleQnA expose opposite limits of local admission. Routing code has the strongest check: execution can reliably reject incorrect programs. However, the SLMs rarely generate a function that passes—only 1 of 8 inputs for the best single model and 2 of 8 across all models—so \name{} escalates 75\% of requests and reaches only the frontier model’s 50\% accuracy. A strong check is therefore useful only when the local models can generate valid candidates. TeleQnA has the opposite problem. Because an incorrect answer is just as well formed as the correct one, the task exposes no intrinsic witness of error. Even the frontier model solves only 84.2\% of the questions, showing that the remaining errors depend on missing knowledge rather than detectable structural violations. Without a task-specific check, \name{} reaches 78.2\% accuracy while escalating 16\% of requests.

\subsection{Improvement over the best single SLM}
We compare the best single SLM with the full \name{} pipeline. On conflict detection, \name{} improves accuracy from 84.0\% to 98.6\%, while on intent translation it improves accuracy from 74.0\% to 93.8\%. The gains vary across the three log-parsing datasets: accuracy increases from 74.3\% to 99.7\% on OpenSSH, from 86.3\% to 91.7\% on HDFS, and from 65.0\% to 87.7\% on Proxifier. On TeleQnA, where no intrinsic check is available, accuracy changes only marginally, from 78.0\% to 78.2\%. These results show that \name{} provides the largest improvements on checkable tasks.

\section{Intrinsic Check as a Boundary}

Our experience with \name{} suggests that the central question for local SLM network automation is not \emph{“which model is strong enough?”} but \emph{“which tasks are checkable enough?”} This reframes local deployment as a property of the task, not just the model. A small model can be useful when the task exposes deterministic structure that makes it easy to rejects bad outputs. The same model is unsafe when the task requires external knowledge or judgment that cannot be checked from the input and local state.

This checkability boundary is a spectrum dependent on strength of the intrinsic checks. At one end are tasks with strong deterministic tests, such as conflict detection with explicit logical contradictions. In the middle are tasks such as intent translation, where grounding and coverage reject common failures but cannot prove policy equivalence. At the weak end are tasks such as log parsing, where invertibility is useful but permissive. Outside the boundary are knowledge-only tasks, where the prompt and candidate answer provide no deterministic way to identify correctness.

\section{Related Work}

\textbf{LLM-driven network automation.}
Recent work has explored LLMs for network automation and networking tasks~\cite{vpp,netconfeval,llm-network-code,netllm,netintent,agenticimr, polysmith, nada}. These systems show that language models can produce useful network artifacts, but they also expose a deployment tension: the strongest results often rely on frontier models, tool-rich workflows, or model-specific adaptations. Our work asks a different question: when can inference remain local without exposing sensitive operator data? \name{} targets that boundary. It uses SLMs as candidate answer generators and uses intrinsic checks to decide which candidates are admissible.

\textbf{Constraints, feedback, and ambiguity in LLM-based networking.}
Verified Prompt Programming combines GPT-4 with syntax and semantic verifiers, plus human input, to synthesize router configurations~\cite{vpp}. Clarify shows that configuration synthesis also requires eliciting ambiguous operator intent, especially when route maps and ACLs overlap in header space and priority cannot be inferred from the prompt alone~\cite{ambiguity}. LeJIT interleaves an SMT solver with generation so model outputs obey domain-specific logic during inference~\cite{lejit}. 

\name{} shares the view that model outputs need external structure, but uses that structure for a different purpose.
Intrinsic checks are cheap deterministic admission tests applied after local generation. A failed check rejects a candidate; a passed check only says the candidate has not violated the tested necessary condition.

\textbf{Network verification, synthesis, and checking.}
Network verification systems check properties such as reachability, isolation, equivalence, policy compliance, and change safety~\cite{netplumber,batfish,minesweeper,netkat, modelfreeverif,llm-code-verification,codet,alphacode,picard}. Network-synthesis systems similarly compile or complete configurations from structured objectives~\cite{propane,netcomplete,aura}. \name{} does not compete with this line as a verification system. It is weaker by design: its checks are per-query necessary-condition tests used to gate local LLM outputs before they reach downstream automation.

\textbf{Model cascades and ensembles.}
Model cascades systems reduce cost by routing queries among models, aggregating generations, or falling back when confidence is low~\cite{frugalgpt,routellm,self-consistency,llm-blender}. We share the generate-and-check philosophy, but use deterministic network checks for a different purpose: to decide when local inference is admissible. A candidate is accepted only when it satisfies external network constraints, not because a model is confident or a majority agrees. Cases that local checks cannot resolve are escalated. Thus, checkability, not ensembling, is the operative boundary.

\section{Conclusion}
Local-first network automation cannot rely on SLM accuracy or model agreement alone. We argue that its viability depends on whether the task exposes cheap, deterministic checks that can reject erroneous outputs. Touchstone instantiates this idea by using SLMs to generate local candidates, admitting only those that survive task-specific intrinsic checks, and escalating the remainder to a frontier LLM. When the checks have low false-accept rates and the local models can generate valid candidates, Touchstone keeps most requests on premises while approaching frontier-model accuracy. Our results support a simple deployment rule: keep inference local when task semantics support precise, low-cost checks; escalate the rest.

\bibliographystyle{ACM-Reference-Format} 
\bibliography{hotnets25-template}

\end{document}